\documentclass[10pt,draft]{dis03}
\usepackage{epsf,amsmath}

\begin{document}

\title{POMERON/ODDERON INTERFERENCE IN DIFFRACTIVE MESON PAIRS PRODUCTION
%\thanks{This work is
%supported by the High-Energy Physics  Foundation}
}

\author{P.~H\"agler \\
 CTP 6-401, LNS, MIT, Cambridge, MA 02139, USA \thanks{Ph.H. thanks the
A.v.H. foundation for the Feodor-Lynen-Fellowship.}\\
        %        indicate your present address (if different from your
         %       normal address), research grant, sponsoring agency, etc.
          %      These are obtained with the {\tt\ttbs thanks} command.},
         B. Pire, \\
         CPhT, {\'E}cole Polytechnique, F-91128 Palaiseau, France
         \thanks{Unit{\'e} mixte C7644 du CNRS},\\
          L.~Szymanowski,\\ So{\l}tan Institute for Nuclear Studies,
         Ho\.za 69, 00-681 Warsaw, Poland, \\  
         \underline{O.V. Teryaev}\thanks{Supported by INTAS
Project 2000-587}, \\
         Bogoliubov Laboratory of Theoretical Physics, JINR, \\ 141980 Dubna, Russia}

\maketitle

\begin{abstract}
\noindent 
We study Pomeron-Odderon interference effects giving rise to charge and single-spin
asymmetries in diffractive electroproduction of a
 $\pi^+\;\pi^-$ pair. The final state interactions and the role of $f_0$
 meson are discussed.

\end{abstract}

Hadronic reactions at low momentum transfer and high energies (for charge-even exchange)
are described in QCD  in terms of
QCD-Pomeron described by the BFKL equation \cite{BFKL}.
The charge-odd exchange is less well understood although the corresponding
BKP equations \cite{BKP} have attracted much attention
recently \cite{Levodd,JW,Vacca1,Korch}, thus
reviving the relevance of phenomenological studies of the Odderon exchange
pointed out years ago in Ref. \cite{LN}.
Unfortunately, 
%the recent studies of specific channels where the
%QCD Odderon contribution is expected
%to be singled out have turned out to be very disappointing.
%In particular, 
recent experimental studies at HERA of
exclusive $\pi^0$ photoproduction  \cite{Olsson}  indicate a very
small cross section for this process which stays in contradiction with
theoretical predictions based on the stochastic vacuum model \cite{Dosh}.

The general feature
of all
meson production processes is that scattering amplitude describing Odderon
exchange enters quadratically in the cross section.
This observation lead to the suggestion
% On
%the one hand, as suggested
in Ref. \cite{Brodsky}, that the study of observables where Odderon effects
are present  at the amplitude level 
%- and not at the squared amplitude
%level - 
is
mandatory to get a convenient sensitivity to a rather small normalization
of this contribution.
This may be achieved by means of  charge asymmetries, for
instance in open charm production \cite{Brodsky}. 
%Since the final state
%quark-antiquark pair has no definite charge parity both Pomeron and Odderon exchanges
%contribute to this process. 
Another example \cite{Nikolaev} is the charge asymmetry in soft
photoproduction of two pions.
Bearing in mind perturbative QCD (pQCD) description, we calculated
the "hard" analogs
of these asymmetries \cite{HPST}, and supplemented them by the
single spin asymmetries,
which may be studied at HERA with polarized lepton beam \cite{HPST1}.
%\section{Method and Results}
We consider the process
$e^-(p_e) N(p_N)  \to  e^-(p_e^{\prime})\pi^+(p_+) \pi^-(p_-)
N^{\prime}(p_N^{\prime})$.  The application of pQCD for the calculation of a
part of this process is justified by the presence of a
 hard scale: the squared mass $-Q^2=-(p_e - p_e^{\prime})^2$
 of the virtual photon,
$Q^2$ being of the order of a few GeV$^2$.
The amplitude of this process 
%(Fig.1)
%\begin{figure}[!thb]
%\vspace*{7.0cm}
%\begin{center}
%\special{psfile=graphs.eps voffset=-60 vscale=40
%hscale= 40 hoffset=10 angle=0}
%\centerline{\epsfxsize=2.9in\epsfbox{kim_mephi_lep.ps}}
%\caption[*]{Contributions of QCD Pomeron and Odderon to 
%meson pair electroproduction}
%\end{center}
%\end{figure}
includes the convolution of a
perturbatively calculable hard subprocess with two non-perturbative inputs,
the 2-pion generalized distribution
amplitude (GDA) and the Pomeron-Odderon (P/O)
proton impact factors. Since
the $\pi^+\pi^-$ system is not a  charge parity
eigenstate, the GDA includes two charge parity components and allows for
a study of the corresponding interference term. The relevant GDA is
just given by the light cone wave function of the two pion system \cite{DGPT}.
The GDAs may be acessed in $\gamma gamma^*$ exclusive production of two 
mesons\cite{DGP}

\section{Observable asymmetries}
We define the  forward-backward or charge asymmetry
$A(Q^{2},t,m_{2\pi }^{2},y,\alpha )$ by
\begin{eqnarray}
\label{ca}
 \frac{\sum\limits_{\lambda =+,-}\int \cos
\theta \,d\sigma (s,Q^{2},t,m_{2\pi }^{2},y,\alpha ,\theta ,\lambda )}{%
\sum\limits_{\lambda =+,-}\int d\sigma (s,Q^{2},t,m_{2\pi }^{2},y,\alpha,\theta ,\lambda )}
%=\frac{\int d\cos
%\theta \cos\theta \;N_{charge}}{\int d\cos
%\theta \;D}
%\label{asym}
\end{eqnarray}
as a weighted integral over polar angle $\theta$ of the relative momentum of
two pions. Although this asymmetry depends on full set of the kinematical
variables, different dependencies, due to factorization, come from different
sources.
The most clean one are the dependencies on $Q^2$, coming
from the hard subprocess, and on dipion mass $m_{2\pi}=\sqrt{(p_+ + p_-)^2}$,
coming from the GDA. The specific form of
$m_{2\pi}$ dependence is explained by the fact, that the phase of the 
GDA\cite{DGPT} should add to the phase shift between Pomeron and Odderon.
%Due to factorization, the same $m_{2\pi}$ dependence should
%define the charge asymmetry at HERMES, providing the possibility to relate
%experiments at rather different energies \cite{HPST1}.
%The $Q^2$ dependence, which is a subject to corrections from
%BFKL, BKP and ERBL evolution, is due to the different $Q^2$ dependence of
%Pomeron and Odderon coefficient functions (perturbative impactfactors),
%the latter having the extra propagators, leading to the decrease of
%asymmetry with $Q^2$. 
%The typical scale of $\alpha_s$ is determined by
%the gluons transverse momenta $\vec{k_i}$ (see Fig.1)
%and we checked that the result is not
%changed substantially if "hard" $\alpha_S(Q^2)$ is used for coefficient
%function and "soft" $\alpha_S \sim 0.5$ is used for non-perturbative
%proton impactfactors.

\begin{figure}[!thb]
\vspace*{7.0cm}
\begin{center}
\includegraphics{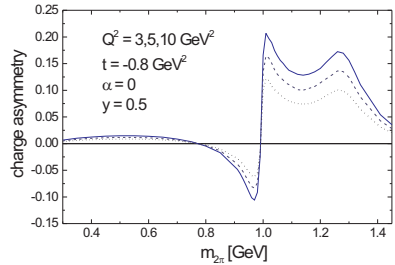}
%\centerline{\epsfxsize=2.9in\epsfbox{kim_mephi_lep.ps}}
\caption[*]{Dependence of charge asymmetry on $m_{2\pi}$ and $Q^2$}
\end{center}
\end{figure}
The latter objects, together with the coefficient functions,
define the dependence of asymmetry
on $t=(p_N-p_N^{\prime})^2$. 
%As to the dependence on the standard
%leptonic variables $y$ (the lepton
%energy fraction carried by the virtual photon) and $\alpha$
%(the angle between lepton and photon scattering planes),
%it is determined by the
%relative size of the virtual photon helicity amplitudes and allows to estimate
%the promising kinematical region \cite{HPST1}.
The single spin asymmetry \cite{HPST1} $A_S(Q^{2},t,m_{2\pi }^{2},y,\alpha )$
contains extra factor $\lambda$ in the numerator of (\ref{ca})
%is defined by
%\begin{eqnarray}
%\label{sa}
% \frac{\sum\limits_{\lambda
%=+,-}\lambda \int \cos \theta \,d\sigma (s,Q^{2},t,m_{2\pi }^{2},y,\alpha
%,\theta ,\lambda )}{\ \sum\limits_{\lambda =+,-}\int d\sigma
%(s,Q^{2},t,m_{2\pi }^{2},y,\alpha ,\theta ,\lambda )}
%\end{eqnarray}
and 
requires to fix the lepton beam polarization $\lambda$.
Contrary to charge asymmetry, this effect is proportional to
the imaginary, rather than to the real part of the interference
term. 
As the Pomeron amplitude is imaginary
and the Odderon one is real
the relative phase between them is the maximal one for the emergence of
single spin asymmetries \cite{OT01}. 
%The effect should be therefore
%maximal for  zero relative phase between isoscalar and isovector
%GDA's, providing
%a complementary probe. This complementarity can be seen
%from the dependence of spin asymmetry on $m_{2\pi}$ and $Q^2$(Fig.2)

%\begin{figure}[!thb]
%\vspace*{7.0cm}
%\begin{center}
%\special{psfile=plotchargeq.eps voffset=-60 vscale=40
%hscale= 40 hoffset=10 angle=0}
%\centerline{\epsfxsize=2.9in\epsfbox{kim_mephi_lep.ps}}
%\caption[*]{Dependence of spin asymmetry on $m_{2\pi}$ and $Q^2$}
%\end{center}
%\end{figure}

%The smaller numerical value of spin asymmetry is to a large extent
%due to the kinematical factor $\sqrt{t}$.
%This asymmetry may be therefore important in the region of large $t$,
%which would become a relevant
%QCD scale and allow to have smaller $Q^2$.

\section{Charge Asymmetry at lower energies and $f_0$ meson contribution} 
The interference effects in electroproduction at lower energies leading 
to the charge asymmetries were already studied both theoretically \cite{LDPS} and experimentally 
\cite{HERMES}. 
The QCD factorization implies, that in this situation the P/O impactfactors should be substituted 
by various Generalized Parton Distributions (GPD), while GDA's are the same as in our case.  
This would result in the same $m_{2\pi}-$dependence of charge asymmetry, 
provided the phases of hard amplitudes are the same, while their difference 
leads to the calculable difference of  $m_{2\pi}-$dependencies. 
The interesting feature of experimental data is the absence of $f_0$-meson contribution, and we would like
to comment on this issue. 
Let us first recall 
that the status of the $f_0$ is very unclear, and
different models are proposed for the $f_0$ structure, where 
the $K \bar K$ molecule plays a special role \cite{Achasov}. 
As the $f_0$ is definitely related to the $K \bar K$ threshold, 
a coupled channel ($K \bar K$ and $\pi \bar \pi$) analysis   is 
obviously needed for the phase shifts analysis above this threshold. 
If $q \bar q$ component of $f_0$ is indeed small, it is quite probable, that 
the $q \bar q$ GDA does not manifest any significant traces of $f_0$ peak.
Taking seriously the absence of $f_0$ signal suggested 
by this reasoning and supported by HERMES 
data  leads to different predictions for our P/O interference effects, 
namely the disappearance of any sizable effect around 950-1000 MeV. The 
signal to search for would then be near the $f_2$ mass.
Moreover, such a connection opens a possibility to study the 
meson structure in the hard processes, focusing on the observables, sensitive to the GDA phases,
like the asymmetries we are considering. 

\section{Conclusions}

We found that a sizable charge asymmetry may be a useful tool
to look for QCD Odderon contribution at HERA.  The spin asymmetry is smaller,
but it can be important at larger $t$ and smaller $Q^2$.
Note finally, that the numerical value of our predictions depend on the adopted
model for proton impact factors, so the observation of the predicted effects
with the different magnitude and/or $t-$dependence might be considered
as their indirect experimental determination. 
%of these important non-perturbative
%objects. 
As a byproduct, the QCD structure of $f_0$ meson may be investigated. 

%\section*{Acknowledgements} We thank all DIS'03 participants
%for their contributions to the DIS'03 Proceedings. This work was
%supported in part by the High Energy Foundation and the World
%Science Agency.

\end{document}